\documentclass[reprint,amsmath,amssymb,aps,prl,showpacs]{revtex4-1}
\usepackage{epsfig}
\usepackage{amsmath}
\usepackage{amssymb}
\usepackage{graphicx}
\usepackage{dcolumn}
\usepackage{hyperref}
\usepackage{bm}
\usepackage{CJK}

\begin{document}

\preprint{APS/123-QED}
\title{Measurement of  specific contact resistivity using scanning voltage probes}
\author{Weigang Wang}
\author{Malcolm R. Beasley}
\email{beasley@stanford.edu}
\affiliation{Geballe Laboratory for Advanced Materials,\\
Stanford University, Stanford, CA 94305
}%
\date{\today}

\begin{abstract}
Specific contact resistivity measurements have conventionally been heavy in both fabrication and simulation/calculation in order to account for complicated geometries and other effects such as parasitic resistance. We propose a simpler geometry to deliver current, and the use of a scanning voltage probe to sense the potential variation along the sample surface, from which the specific contact resistivity can be straightforwardly deduced. We demonstrate an analytical example in the case where both materials are thin films. Experimental data with a scanning Kelvin probe measurement on graphene from the literature corroborates our model calculation.
\end{abstract}

\maketitle

The contact between any two conductors exhibits an interface resistance.  The intensive quantity that characterizes this contact resistance is the specific contact resistivity, defined as the ratio of the voltage across the interface and the current density through the interface. The accurate measurement of this quantity is of great technical and scientific importance\cite{Murakami1998,Zhang2003,Deen2006}. In this {\it Letter} we propose a method of measuring specific contact resistivity based on the use of scanning voltage probes.  This approach is simpler than traditional methods with the added benefit that inhomgeneities of the specific contact resistivity can be detected.

The methods that have hitherto been developed include, e.g., the transfer length measurement (TLM) and the cross bridge Kelvin resistor (CBKR) method. Both methods require accurate fabrications of test structures. The TLM method\cite{Schroder2006} extrapolates from two point resistance measurements of devices with different lengths to obtain the contribution of contact resistance in the total resistance. Accuracy in fabrication and homogeneity are required for this method. In the CBKR method\cite{Schroder2006,Loh1985,Stavitski2009}, a three dimensional test structure is fabricated and the contact resistance measured in a four point configuration. The measurement result then undergoes a self-consistent algorithm to obtain the specific contact resistivity. A major issue  when applying these methods is the interplay of the length scales  of the test structure geometry and the materials dependent current transfer length\cite{Schroder2006}.

\begin{figure}
\begin{center}
\includegraphics[width=3.2in,bb=0 0 230 100]{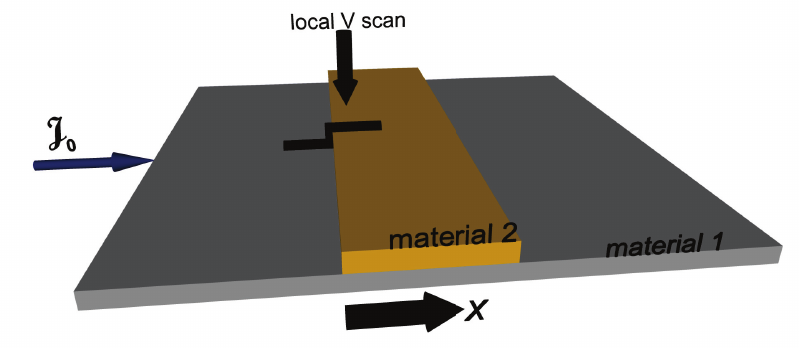}
\caption{Schematic of proposed measurement of specific contact resistivity between materials 1 and 2. A planar structure should be fabricated to present a well defined boundary between the two materials. Local voltage variation due to the existence of a sheet current density $\mathcal{J}_0$ across the boundary can be measured and from the measured profile the specific contact resistivity can be calculated. A jump in the voltage when scanned from above is expected due to the specific contact resistivity.}\label{device}
\end{center}
\end{figure}

In our approach we utilize information from the spatial variation of the local voltage in the simple structure depicted in Fig. \ref{device} to determine the specific contact resistivity. In this structure there is essentially no geometric length scale, and the current transfer length is  determined from the scanned voltage profile. Both material 1 and material 2 can be either thin or thick, although a thin film might be appropriate for fabrication purposes and easier to model.  The spatial variation of the voltage across the boundary of the two materials can be measured, and the data fitted to obtain the value of specific contact resistivity. Also, the voltage variation along both materials can be used to determine the (sheet) resistance of both materials.

Various instruments can be used to measure the local voltage variation along the sample surface.  These include scanning potentiometry and scanning electrostatic voltage probes.  For scanning potentiometry, the natural choices are scanning tunneling potentiometry (STP)\cite{Rozler2008,Wang2012}, which has the highest spatial resolution ($<$10nm), and scanning conducting AFM, which is perhaps simpler but with courser spatial resolution\cite{Hersam1998,Nakamura2005}.  For electrostatic probes, scanning Kelvin probes \cite{Puntambekar2003}, and scanning single electron transistors\cite{Martin2008} (which is a form of Kelvin probe with high sensitivity) are suitable choices.  While both types of probes will work, scanning potentiometry has the attractive feature that it measures the transport potential directly whereas electrostatic probes have the complication of dealing with spatial variations of the electrostatic potential associated with the patch effect.

The calculational description of the test structure is well developed\cite{Mohney2005,Schroder2006}. In our case it is easier due to the simpler structure. In the case where both films are thin (the condition for a film to be thin will be specified), and sheet resistance and sheet current density can be used for both materials, we write down the analytical solution below as a demonstration on how this method works.

%\begin{figure}
%\begin{center}
%\includegraphics[width=3.6in,bb=0 0 1024.42 489.08]{J.pdf}
%\caption{Numerical examples of sheet current density redistribution near the boundary between two materials as calculated in equation (\ref{solution}). The sheet current density is normalized to $\mathcal{J}_0$, the total sheet current density applied; and $x$-direction displacement is normalized to $l_1\equiv\sqrt{\mathcal{R}_s/R_{\square1}}$.}\label{Jprofile}
%\end{center}
%\end{figure}

\begin{figure}
\begin{center}
\includegraphics[width=3.6in,bb=0 0 259 247]{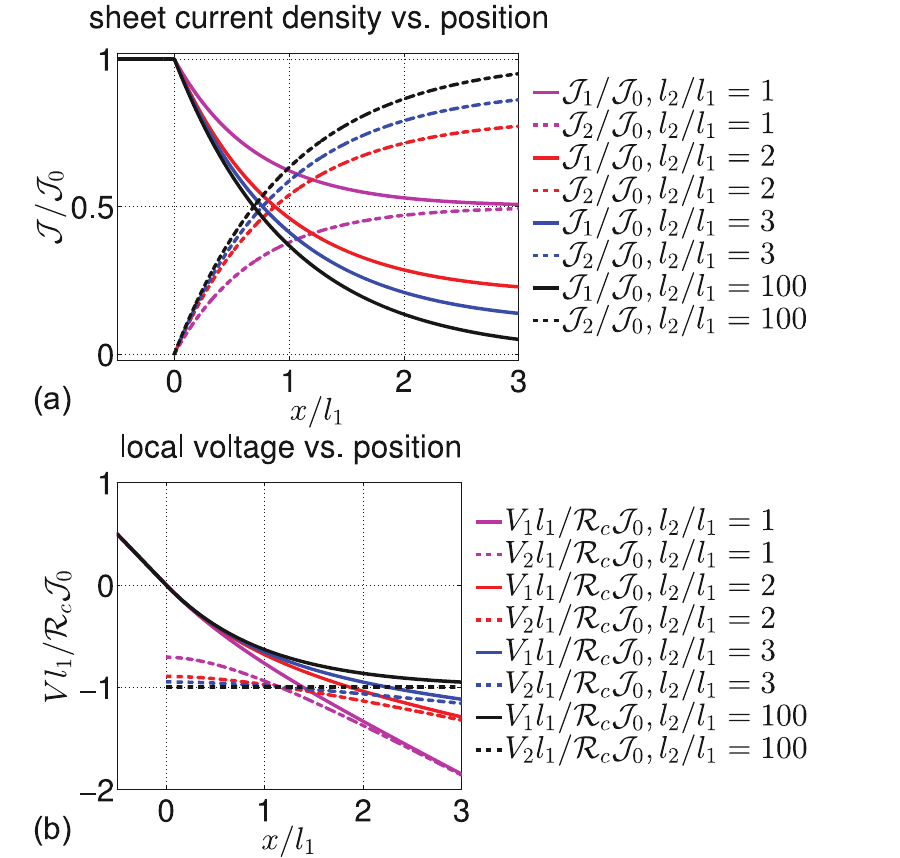}
\caption{Numerical examples of solution (\ref{solution}). (a) Sheet current density redistribution near the boundary between two materials as calculated in equation (\ref{solution}). The sheet current density is normalized to $\mathcal{J}_0$, the total sheet current density applied; and $x$-direction displacement is normalized to $l_1\equiv\sqrt{\mathcal{R}_c/R_{\square1}}$. (b) Local voltage variation near the boundary between two materials as calculated in equation (\ref{solution}). The voltage $V(x)$ is normalized to $\mathcal{J}_0\mathcal{R}_c/l_1$; and $x$-direction displacement is normalized to $l_1\equiv\sqrt{\mathcal{R}_c/R_{\square1}}$. Note that if scanned from the top in Fig. \ref{device}, the scanning probe picks up $V_1$ at $x<0$ but $V_2$ at $x>0$. Hence a jump in the measured local voltage will be observed.}\label{JV}
\end{center}
\end{figure}

%\begin{figure}
%\begin{center}
%\includegraphics[width=3.6in,bb=0 0 1024.42 489.08]{V.pdf}
%\caption{Numerical examples of local voltage variation near the boundary between two materials as calculated in equation (\ref{solution}). The voltage $V(x)$ is normalized to $\mathcal{J}_0\mathcal{R}_s/l_1$; and $x$-direction displacement is normalized to $l_1\equiv\sqrt{\mathcal{R}_s/R_{\square1}}$. Note that if scanned from the top in Fig. \ref{device}, the scanning probe picks up $V_1$ at $x<0$ but $V_2$ at $x>0$. Hence a jump in the measured local voltage will always be observed.}\label{Vprofile}
%\end{center}
%\end{figure}

The following calculation assumes that the films are thin for both material 1 and material 2, described by a sheet resistance $R_{\square1,2}\equiv\rho_{1,2}/d_{1,2}$, where $d_{1,2}$ are the individual thicknesses of the two films. The specific contact resistivity between the two materials is $\mathcal{R}_c$. The sheet current densities in the two materials are $\mathcal{J}_{1,2}$, where $\mathcal{J}_{2}$ is only valid to the right of the boundary ($x>0$). Assume that the electrode is macroscopic in size, i.e., the other boundary is infinitely far from the boundary near which the scan is performed. To the left of the boundary ($x<0$), due to current conservation the sheet current density is a constant:
\begin{equation}
\mathcal{J}_{1}(x)=\mathcal{J}_{0}\quad x<0\ ,
\end{equation}
and equals the sheet current density applied to the device. Thus, to the left of the boundary the local voltage is:
\begin{equation}
V_1(x)=-R_{\square1}\mathcal{J}_{0}x\quad x<0\ .
\end{equation}

To the right of the boundary ($x>0$), the differential equations are:
\begin{equation*}
-\dfrac{dV_1(x)}{dx}=\mathcal{J}_{1}(x)R_{\square1}\quad\mbox{(material 1)},
\end{equation*}
\begin{equation*}
-\dfrac{dV_2(x)}{dx}=\mathcal{J}_{2}(x)R_{\square2}\quad\mbox{(material 2)},
\end{equation*}
\begin{equation*}
\mathcal{J}_{1}(x)+\mathcal{J}_{2}(x)=\mathcal{J}_{0}\quad\mbox{(current conservation)},
\end{equation*}
\begin{equation*}
J_c=\dfrac{V_1(x)-V_2(x)}{\mathcal{R}_c}\quad\mbox{(material interface)},
\end{equation*}
\begin{equation}\label{ode}
J_c=-\dfrac{d\mathcal{J}_{1}(x)}{dx}\quad\mbox{(current conservation at interface)},
\end{equation}
with the boundary conditions:
\begin{equation}\label{boundary}
V_1(0)=0\ ,\quad \mathcal{J}_{1}(0)=\mathcal{J}_{0}\ .
\end{equation}

The above equations can be analytically solved. For convenience the following two parameters are used: $l_{1,2}=\sqrt{\mathcal{R}_c/R_{\square1,2}}$, and we denote $l\equiv l_1l_2/\sqrt{l_1^2+l_2^2}$. These are the transfer lengths in respective materials\cite{Schroder2006}. The solution for $x>0$ then reads:
\begin{equation*}
\mathcal{J}_{1}(x)=\mathcal{J}_{0}\left(\dfrac{l^2}{l_1^2}e^{-x/l}+\dfrac{l^2}{l_2^2}\right)\ ,
\end{equation*}
\begin{equation*}
\mathcal{J}_{2}(x)=\mathcal{J}_{0}\cdot\dfrac{l^2}{l_1^2}\left(1-e^{-x/l}\right)\ ,
\end{equation*}
\begin{equation*}
V_1(x)=-\mathcal{R}_c\mathcal{J}_{0}\left[\dfrac{l^3}{l_1^4}\left(1-e^{-x/l}\right)+\dfrac{l^2}{l_1^2l_2^2}\cdot x\right]\ ,
\end{equation*}
\begin{equation}\label{solution}
V_2(x)=-\mathcal{R}_c\mathcal{J}_{0}\left(\dfrac{l^3}{l_1^4}+\dfrac{l^3}{l_1^2l_2^2}e^{-x/l}+\dfrac{l^2}{l_1^2l_2^2}\cdot x\right)\ .
\end{equation}

Some numerical examples of the above solution, also including $\mathcal{J}_{1}$ and $V_1$ at $x<0$, are presented in Fig. \ref{JV}. The curves show how part of the sheet current density is transferred from material 1 to material 2 in the existence of contact resistance. The current transfer is the largest at the boundary, and exponentially decays with a length scale of $l$. As a result there is a voltage difference between the two materials at the boundary, which gradually becomes smaller over a length scale of $l$ as sheet current is transferred.

The solution self-scales, as is evident from both the form of the formula and the numerical examples. The major features of the solution are: a slope at $x<0$ of value:
\begin{equation}
\beta_1=-\mathcal{R}_c\mathcal{J}_{0}/l_1^2=-R_{\square1}\mathcal{J}_{0}\ ,
\end{equation}
an asymptotic slope at $x\gg0$ of value:
\begin{equation}
\beta_2=-\mathcal{R}_c\mathcal{J}_{0}l^2/(l_1^2l_2^2)=-R_{\square1}\mathcal{J}_{0}l^2/l_2^2\ ,
\end{equation}
and a jump in the local voltage variation, if measured from the top of the device as depicted in Fig. \ref{device}, at the boundary $x=0$, of value:
\begin{equation}
\Delta V=-\alpha\mathcal{R}_c\mathcal{J}_{0}/l_1=-\alpha R_{\square1}\mathcal{J}_{0}\cdot l_1\ .
\end{equation}
Here, $\alpha=l/l_1$ is a numerical factor that depends on the ratio of $l_1$ and $l_2$, as can be seen in the numerical examples in Fig. \ref{JV}(b).

The above scaling provides a very convenient procedure to determine experimentally all three material dependent parameters: $R_{\square1}$, $R_{\square2}$, and $\mathcal{R}_c$, from one scanned local voltage curve across the boundary. 1) From the known applied current to the device, $\mathcal{J}_{0}$ can be calculated, and $R_{\square1}=|\beta_1|/\mathcal{J}_{0}$ can be obtained. 2) From the ratio of $\beta_1/\beta_2$, $R_{\square2}=R_{\square1}\dfrac{|\beta_2|}{|\beta_1|-|\beta_2|}$ can be obtained. 3) From the size of the jump, $l_1=|\Delta V|/(\alpha|\beta_1|)$ can be obtained, from which $\mathcal{R}_c=R_{\square1}l_1^2$ can be calculated. If necessary, curve fitting can also be employed to improve quality of calculated results.

We note the experimental result shown in Fig. 4(d) in reference \cite{Yu2009} , in which a scanning Kelvin-probe measurement was performed on a device made of graphene (material 1) and Cr/Au (material 2). The authors of that paper correctly extracted the contribution of contact resistance from their data. We would point out, however, that this set of data can also be used to estimate the specific contact resistivity value. From an estimated $R_\square=200\Omega/\square$ for graphene and $\Delta V/|\beta_1|=2\mu$m from their data, we estimate the specific contact resistivity to be $\mathcal{R}_c\approx1\times10^{-5}\Omega\cdot\mbox{cm}^2$, which is very close to values obtained in the literature\cite{Venugopal2010}.

It can also be seen that $l$ is a characteristic scale in the local voltage variation and sheet current density distribution. Thus it is clear that the film thicknesses should be much smaller than $l$ in order for the set of equations (\ref{ode}) and (\ref{boundary}) to apply. When film thicknesses are larger than or comparable to $l$, a two-dimensional calculation is needed to account for current crowding at the edges of the electrodes. Nevertheless, the main features -- the two (asymptotic) slopes and one jump at the boundary should be the same. The difference would be a different numerical value for $\alpha$ than $l/l_1$. One exception would be the case $l_2\gg l_1$, i.e. a much more conductive upper material than lower material. In this case, as long as the lower material is thin with $d_1\ll l_1$, the upper material is an equipotential, and the thickness $d_2$ does not matter.

\begin{figure}
\begin{center}
\includegraphics[width=3.6in,bb=0 0 259 225]{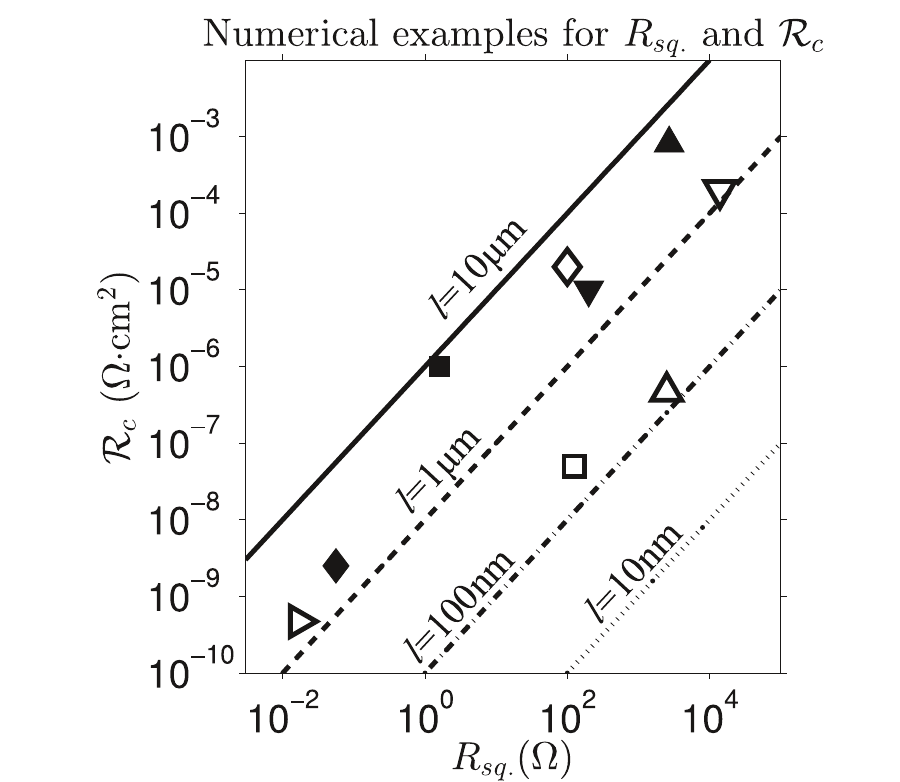}
\caption{Numerical values of $\mathcal{R}_c$ and $R_{\square}$ for different materials. The lines indicate constant transfer lengths. The markers are respective values for: $\square$ NiSi or PtSi to p or n doped Si\cite{Stavitski2006}; $\blacksquare$ Si/Ti/TiN to n doped Ge\cite{Martens2011}; $\vartriangle$ Ti/Al/Mo/Au to 100nm n doped GaN\cite{Kumar2002}; $\blacktriangle$ Ni/Au to GaN nanowire\cite{Stern2006}; $\triangledown$ Ti/Au to 1$\mu$m Al doped ZnO\cite{Kim2000}; $\blacktriangledown$ Cr/Au to graphene\cite{Yu2009}; $\lozenge$ Ni to n doped 3C-SiC\cite{Wan2002}; $\blacklozenge$ 10nm Al to 10nm W\cite{Kang1988}; and $\vartriangleright$ 10nm Cu to 10nm Cu\cite{Gueguen2010}.}\label{examples}
\end{center}
\end{figure}

To help  assess the utility of our approach to the determination of the specific contact resistivity, we plot representative values of $R_{\square}$ and $\mathcal{R}_c$ in Fig. \ref{examples}.  The diagonal lines in the figure indicate contours of constant current transfer length. As seen in the figure,  the values of transfer lengths are between 100nm and 10$\mu$m. This range is comfortably above the spatial resolution of instruments mentioned above.  Note also that as the specific contact resistivity decreases the high spatial resolution of modern scanning probes become more and more essential.

Practical devices have finite roughnesses ranging from nanometers to tens of nanometers\cite{Kim2000,Kumar2002,Nakayama2004}. Even though the roughnesses quoted here are small compared to transfer length values presented in Fig. \ref{examples}, when the roughness is high, there can be a systematic error due to the larger contact area than the nominal value, i.e., the measured specific contact resistivity is proportionally smaller than the true value if the contact region is rough. This can be true for both the conventional methods and our proposed method. The measured specific contact resistivity is an effective value not accounting for the roughness of the interface in both cases. We note that some of the proposed instruments in our method, e.g., scanning conducting AFM or STP, have the capability of measuring topographical roughnesses on both materials and can thus provide an estimate of the roughnesses, though this is not the true roughness of the interface. Systems with smooth surfaces and interfaces such as \cite{Arulkumaran2013,Lee2011} are desirable for both contact resistivity measurement and fabrication scale down.

Another finite contribution to the specific contact resistivity due to sample roughness is the effect of non-uniform current distributions across the interface between two materials, across the boundaries between two materials, and within the two materials. If the specific contact resistivity is small and the transfer length scales are comparable to the surface roughnesses, the situation is more complicated. These effects will clearly affect the accurate determination of specific contact resistivity in both the conventional methods and in our method, and a simulation is required to address the issues. We stress that a scanning voltage probe measurement in our method can be used to investigate the deviation from ideal curves in the one-dimensional theory, and the two-dimensional potential maps on the same sample can provide an empirical estimate to the roughnesses. These estimates may also help addressing the systematic error discussed in the previous paragraph. Such simulations required to quantify the effects of roughness will be pursued in a separate paper.

Another interesting limiting case is when the sheet resistances of both materials are zero $R_{\square1,2}=0$, i.e., superconducting. If the sheet current density $\mathcal{J}_0$ flows through only one material, the case is trivial, namely no voltage drop across this material. However, if the sheet current is forced into one material and out of the other, a voltage drop is expected: superconductivity dictates that both materials are equipotentials; because of the current transfer, a voltage drop exists at the interface of the two materials. If the specific contact resistivity is homogeneous, the current transfer has to be homogeneous over the interface, consistent with $l=\infty$. Hence in this case a four point measurement across the interface and a good measurement of the interface surface is sufficient in determining the specific contact resistivity $\mathcal{R}_c=\Delta V\cdot A/I$, where $A$ is the area of the interface and $I$ is the current through the interface.

We also note that the problem of thermal boundary resistance in the same geometry as Fig. \ref{device} has a solution of the same form with our solution (\ref{solution}). Hence a scanning thermometer across the boundary of a thermal interface might provide a similarly convenient measurement of specific thermal boundary resistivity.

In summary, the method we propose requires only a simple electrode-sample geometry and at the most a light numerical simulation. The benefit of self-calibration provides high reliability. Multiple instruments are available to perform this measurement.

\begin{acknowledgements}
We would like to thank David Goldhaber-Gordon and Francois Amet for helpful discussions. Support for this work came from the Air Force Office of Scientific Research MURI Contract \# FA9550-09-1-0583-P00006. One of us (WW) further acknowledges the generous support of a Stanford Graduate Fellowship. We thank an anonymous referee for pointing out the issue of interface roughness.
\end{acknowledgements}

%\bibliography{contactBib}

%merlin.mbs apsrev4-1.bst 2010-07-25 4.21a (PWD, AO, DPC) hacked
%Control: key (0)
%Control: author (8) initials jnrlst
%Control: editor formatted (1) identically to author
%Control: production of article title (-1) disabled
%Control: page (0) single
%Control: year (1) truncated
%Control: production of eprint (0) enabled
%

\end{document}